\documentclass[10pt, letterpaper, reprint, superscriptaddress, aps, prl]
{revtex4-1} \usepackage[latin1]{inputenc} 
\usepackage{amsmath}
\usepackage{amsfonts} \usepackage{amssymb} \usepackage{float}
\usepackage{graphicx} \usepackage{soul} \usepackage[mathscr]{eucal}
\usepackage{dcolumn}
\usepackage{bm}
\usepackage{multirow}
\usepackage{array}
\usepackage{tabularx}
\usepackage{bm}

\begin{document}


\title{Colloidomers: freely-jointed polymers made of droplets}
\author{Angus McMullen}
 \affiliation{Physics Department, New York University, New York, New York 10003, USA}
\author{Miranda Holmes-Cerfon}
 \affiliation{Courant Institute of Mathematical Sciences, New York University, New York, New York 10003, USA}
 \author{Francesco Sciortino}
 \affiliation{Dipartimento di Fisica, Sapienza Universita' di Roma, Rome, Italy}
 \author{Alexander Y. Grosberg}
 \affiliation{Physics Department, New York University, New York, New York 10003, USA}
\author{Jasna Brujic}
 \affiliation{Physics Department, New York University, New York, New York 10003, USA}

\begin{abstract}
An important goal of self-assembly is to achieve a preprogrammed structure with high fidelity. Here, we control the valence of DNA-functionalized emulsions to make linear and branched model polymers, or `colloidomers'. The distribution of cluster sizes is consistent with a polymerization process in which the droplets achieve their prescribed valence. Conformational dynamics reveals that the chains are freely-jointed, such that the end-to-end length scales with the number of bonds $N$ as $N^{\nu}$, where $\nu\approx3/4$, in agreement with the Flory theory in 2D. The chain diffusion coefficient $D$ approximately scales as $D\propto N^{-\nu}$, as predicted by the Zimm model. Unlike molecular polymers, colloidomers can be repeatedly assembled and disassembled under temperature cycling, allowing for reconfigurable, responsive matter.     
\end{abstract}

\maketitle
Experimental models of molecular polymers on the colloidal~\cite{Feng2013, Zhang2017, Zerrouki2008, Sacanna2010, Biswas2017, Dreyfus2005, coluzza2013design} or granular length scales \cite{Zou2009,Tricard2012, Rozynek2017, Miskin2013, vutukuri2012colloidal, Miskin2014, Hoy2017} have been proposed in the literature. These studies have demonstrated how particle specificity, shape anisotropy, or non-equilibrium interactions can lead to the formation of chains. Assembly occurs either through thermal agitation, or the particles are aligned by external drives, such as gravity, capillarity, shaking, or magnetic and electric fields~\cite{Song2015, Rozynek2017}. Despite these innovative proofs-of-concept, it remains an experimental challenge to design particles that self-assemble into flexible chains in bulk, leading to analogues of polymer solutions on different length scales.  

	Here we demonstrate a promising avenue for self-assembly, in which thermal activation leads to the polymerization of DNA-coated droplet monomers into chains, i.e. `colloidomerization'. The process of chain growth follows random aggregation, such that the average weight of the colloidomers and its statistical distribution can be predicted theoretically. These colloidal polymers allow for the visualization of micron-sized monomers, a length scale that is inaccessible in molecular systems. Having access to the conformational dynamics of a large pool of chains allows us to study their statistical physics. 
    
    While polymer theories have been validated at the single-molecule level using DNA imaging and force spectroscopy \cite{Smith1996,Bustamante1994,Maier1999}, our experimental model system examines whether these theories hold on colloidal length scales. The most important advantage of using droplets instead of solid particles as the monomer units is that the DNA bonds between them are fully mobile along the droplet surface. The particles can therefore rearrange after binding and dynamically explore their equilibrium structures on a timescale fixed by the droplet diffusion constant. Therefore, these colloidomers behave like freely-jointed polymers. By contrast, solid colloidal polymers have limited flexibility~\cite{Shah2014,Song2015,Sacanna2010, Biswas2017}, while membrane-coated colloids can have preferred bond angles, restricting their relative motion~\cite{chakraborty2017colloidal}. 
    
    \begin{figure}[!b]%
     \centering	\includegraphics{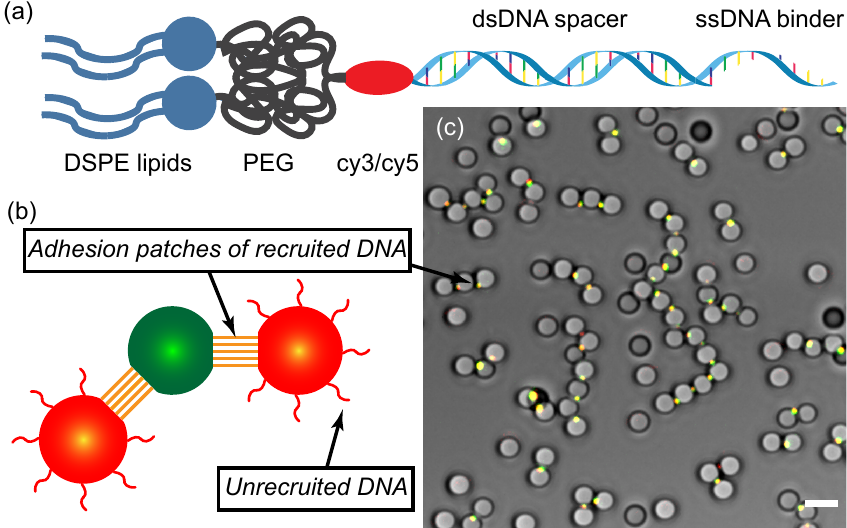}%
     \caption{(a) A schematic of the DNA-lipid construct grafted on the droplet. (b) Mixing emulsions with complementary DNA strands leads to their binding to form a trimer. (c) A bright-field image overlayed with a fluorescence image shows adhesions in self-assembled droplet chains.  Scale bar is 5\,$\mu$m.}
     \label{fig:intro}
 \end{figure}

    We find excellent agreement between the data and the scaling exponents predicted by the Flory theory of 2D freely-jointed chains with excluded volume, down to a chain size of only a handful of droplets. The self-diffusion coefficient of the colloidomers follows the Zimm scaling with length, highlighting the importance of hydrodynamic interactions.
    
    One important difference between colloidomers and molecular polymers is that their particulate nature allows us to control monomer interactions to trigger chain assembly, disassembly, and reconfiguration. For example, here we show that cycling temperature to melt and reassemble the DNA bonds between the droplets can produce statistically similar distributions of chains. Previously, we have shown that the linear sequence of droplets with different DNA flavors can be programmed via DNA toe-hold displacement reactions~\cite{Zhang2017}. In the future, these chains will serve as backbones for colloidal folding into complex 3D structures via secondary interactions, as proposed in~\cite{Cademartiri2015, Tricard2012}.
    
    To achieve a good yield and high fidelity of chains, we synthesize thermal droplets with a uniform coverage of sticky DNA, ensuring both fast dynamics and valence control. We make monodisperse PDMS droplets using a method adapted from~\cite{Elbers2015} and outlined in detail in~\cite{Zhang2017}. After we synthesize the droplets, we incubate them with a variable amount of binding DNA. As shown in Fig.\ref{fig:intro}(a), the DNA binders are comprised of a pseudo-random spacer of 50 bases, followed by a binding sequence of 20 bases, with one complementary strand labeled with a Cy3 and the other with a Cy5 fluorophore (see~\cite{SM} for DNA sequences).  This molecule is then reacted with a lipid (DSPE-PEG2000-DBCO, Avanti) through a copper-free click reaction. The hydrophobic part of the lipid anchors the DNA strands to the surface of the droplet. We react an additional strand, complementary to the random spacer prior to the sticky end, with another lipid molecule. This DNA-lipid complex hybridizes with that of the spacer of the binding DNA strand, anchoring the entire complex via two lipid molecules instead of one. This prevents the migration of DNA from droplet to droplet. The droplet buoyancy pins them close to the glass surface of the sample cell, confining the dynamics to two dimensions.
    
 \begin{figure}[!tb]%
     \centering	\includegraphics{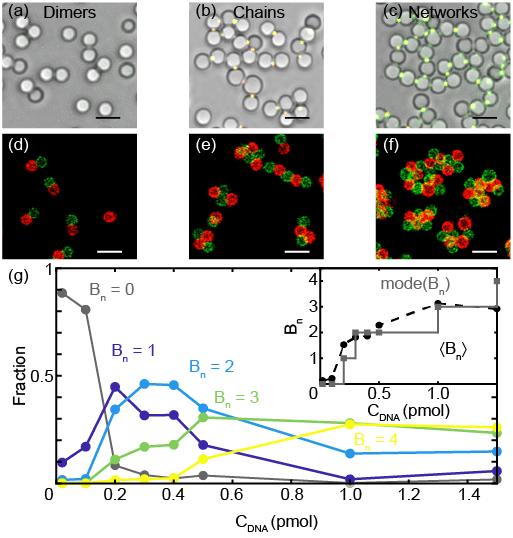}%
     \caption{Bright-field ((a), (b), (c))   and fluorescence ((d), (e), (f), see~\cite{SM} for details) images of droplet structures show an increase in valence as a function of $C_{DNA}$.  Scale bars are 5\,$\mu$m. (g) The number of bonds per droplet $B_n$ increases, as shown by the average and the mode in the inset.}
     \label{fig:valence}
 \end{figure}
    
    A droplet functionalized with single-stranded DNA binds to a complementary strand on another droplet{~\cite{Hadorn2012, Pontani2012, Feng2013, Angioletti-Uberti2014} through DNA hybridization of the single-stranded (sticky) ends, as shown in Fig.\ref{fig:intro}(b). The choice of two distinctly functionalized droplets to make a heterocopolymer (instead of a homopolymer) plays also a crucial role in decreasing, if not suppressing, aggregates with minimal closed loops. The adhesion patch has a much higher density of DNA than on the perimeter of the droplet, as shown by the co-localized fluorescent signal in the vicinity of droplet contacts in Fig.\,\ref{fig:intro}(c). Since the density of DNA that can be recruited into a single patch is limited by geometry, any remaining DNA on the surface of the droplet can form a second patch with another droplet. Starting with a 1:1 mixture of complementary droplets at an area fraction\,$\approx0.2$, we observe that the reaction rate of droplet-droplet binding increases linearly with the bulk DNA concentration up to $C_{DNA}=2$\,pmol, above which the rate plateaus~\cite{SM}, as expected. Therefore, the assembly process reaches a steady state after a week for low $C_{DNA}$ down to a day for high $C_{DNA}$.    
    
    Empirically, at steady state we observe an increase in droplet valence as a function of $C_{DNA}$, from dimers in Fig.\,\ref{fig:valence}(a,d), through chains in (b,e), to cross-linked networks in (c,f), resembling fibrous gels~\cite{Janmey2007}. For a given $C_{DNA}$, we track the droplets in 2D over time~\cite{Crocker1996} to determine the evolution of the droplet bond network. We consider two droplets bound only if their bond persists over one minute, consistent with the fact that DNA bonds are practically irreversible at room temperature. 
    
     Figure\,\ref{fig:valence}(g) shows the fraction of droplets with a given bond number $B_n$ at steady state, as a function of $C_{DNA}$. The spread in $B_n$ at a given $C_{DNA}$ may arise from nonuniform DNA coverage between droplets, any residual polydispersity, or kinetic or steric bottlenecks that prevent the droplets from reaching their prescribed valence. The inset in Fig.\ref{fig:valence}(d) shows that the average $B_n$ increases smoothly with concentration, while the mode (the most probable $B_n$ value) exhibits a stepwise increase, allowing us to tune the self-assembly.  
    
 \begin{figure}[!tb]%
     \centering	\includegraphics{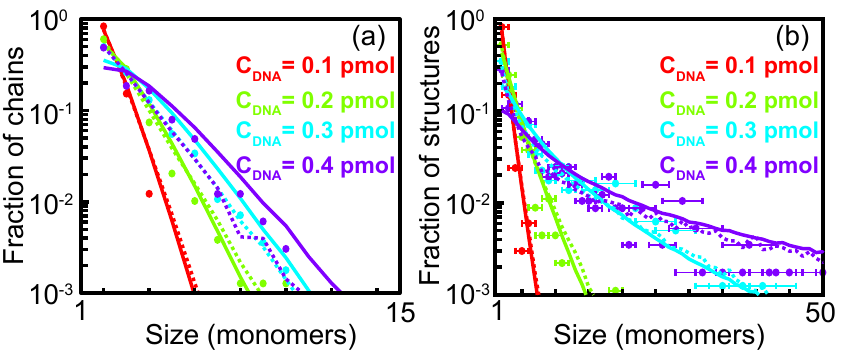}%
     \caption{(a) Distribution of structure sizes for linear chains and all clusters (b) for the indicated given  loadings. The solid lines are predictions of the random branching model (RBM), while dashed lines are the results of a Monte Carlo simulation. The horizontal error bars in (b) indicate monomer bin size.}
     \label{fig:distribution}
 \end{figure}
 
    Optimizing for valence two at $C_{DNA} = 0.2$\,pmol, the droplets assemble into linear or branched polymers in $63\%$ of structures, while $28\%$ of droplets remain unreacted monomers, and $9\%$ constitute aggregates with higher valencies. The distribution of cluster sizes is shown by the distributions in Fig.\,\ref{fig:distribution}(a) for the subset of clusters that are in linear chains and in Fig.\,\ref{fig:distribution}(b) for all structures. 
We compare these observed distributions to both a random branching model and to Monte Carlo simulations,
assuming in both cases the experimentally measured valence distribution as an input. The random branching model~\cite{Grimmett2001, SM} computes the size distribution of colloidomers assuming that they do not form loops, that monomers attach independently, and that each colloidomer is saturated, with no unused bonds~\cite{SM}. Monte Carlo simulations mimic the experiment following the procedure outlined in \cite{SM}. 

The close agreement between the experimental results, the random branching model, and the Monte Carlo simulations, implies that the self-assembly is a random process, which goes to completion with minimal steric or kinetic inhibition. Even though the bonds are irreversible at room temperature, the subset of linear chains and cluster sizes is reproduced with the simple assumption of the same input valence. This result lends support to the hypothesis that each droplet has a valence prescribed only by the concentration of DNA molecules on its surface. Fluctuations in the amount of DNA arise from the stochastic process associated with loading the DNA on the droplets. While the data and the simulations are fully consistent with an exponential distribution, the branching model predicts a deviation for short chains. The characteristic length scale is an increasing function of $C_{DNA}$, offering the possibility to generate long chains and test the scaling relations of flexible polymers.

 \begin{figure}[!tb]%
     \centering	\includegraphics{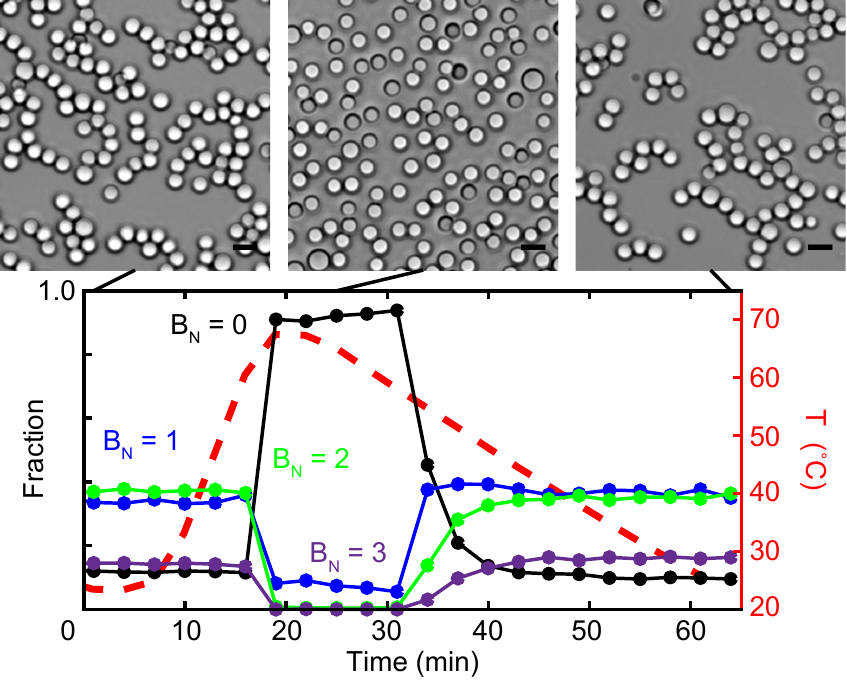}%
     \caption{Evolution of the bond number distribution of a sample as a function of time under the temperature ramp shown in red. The corresponding images of the droplet structures are shown above. Scale bar is 5 $\mu$m.}
     \label{fig:melting}
 \end{figure}

In contrast to molecular polymers, colloidomers can be disassembled upon heating above the DNA melting temperature and reassembled into statistically similar chains upon cooling, as shown by the reversible valence distribution in Fig.\ref{fig:melting}. The exponential decay in the fraction of monomers (black line) in response to the cooling temperature ramp reveals a rate of droplet binding of 1/7\,min, such that the $B_N$ distribution reaches a steady state after only 20 minutes. The fast reassembly is due to the presence of pluronic surfactant (F68) in the system. 
At high temperatures, the surfactant causes a weak depletion between droplets, increasing the assembly rate far beyond that observed at room temperature~\cite{SM}. 
This evidence of reversibility in the droplet valence is in agreement with the numerical prediction that aggregation (aging) dynamics is similar to that occurring in equilibrium for systems with a low valence~\cite{Ruzicka2011, corezzi2012chemical}. The possibility to manipulate droplet-droplet bonds by temperature or using specific DNA reactions \cite{Zhang2009} presents a useful tool to configure colloidomer solutions {\it in situ}.

Next, we investigate the conformational statistics of trimers in 2D to test whether the droplet-droplet bonds are freely-jointed. 
The bond angle subtended by the centers of the two end-cap droplets on the center of the middle droplet is shown in the inset in Fig.\,\ref{fig:joint}(b). 
Apart from the angle $\frac{\pi}{3}$ excluded on both sides of the center droplet, the histogram in the Fig\,\ref{fig:joint}(a) shows that droplets explore bond angles with a flat probability.
To characterize the angular displacement $\Delta\theta(t)$, we fit $-\ln \left< \cos \Delta\theta (t) \right> $, which grows as $D_{rot} t$ with time, where $D_{rot}$ is the rotational diffusion coefficient (see~\cite{SM}).
This expression accounts for the bounded nature of the angles. 
This generalized law for angular diffusion reduces to the law for translational diffusion, $\left< \Delta\theta^2 \right> \simeq 2 D_{rot} t$, for small $\Delta \theta$.
 
\begin{figure}[!tb]%
     \centering	\includegraphics{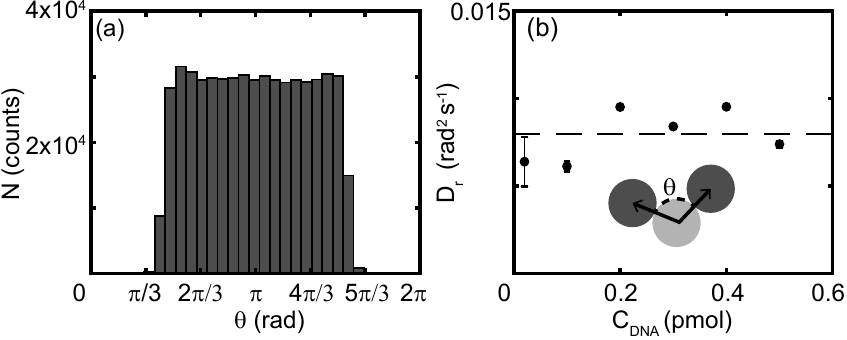}%
     \caption{(a) A histogram of the explored angular configurations, corrected for out of plane motion. Angles above $\frac{\pi}{3}$ and below $\frac{5\pi}{3}$ are allowed. (b) The measured angular diffusion coefficient for the bond reorganization as a function of the DNA loading density. There is no correlation between DNA loading density and bond viscosity.}
     \label{fig:joint}
 \end{figure}
 
  \begin{figure*}[t]%
     \centering	\includegraphics{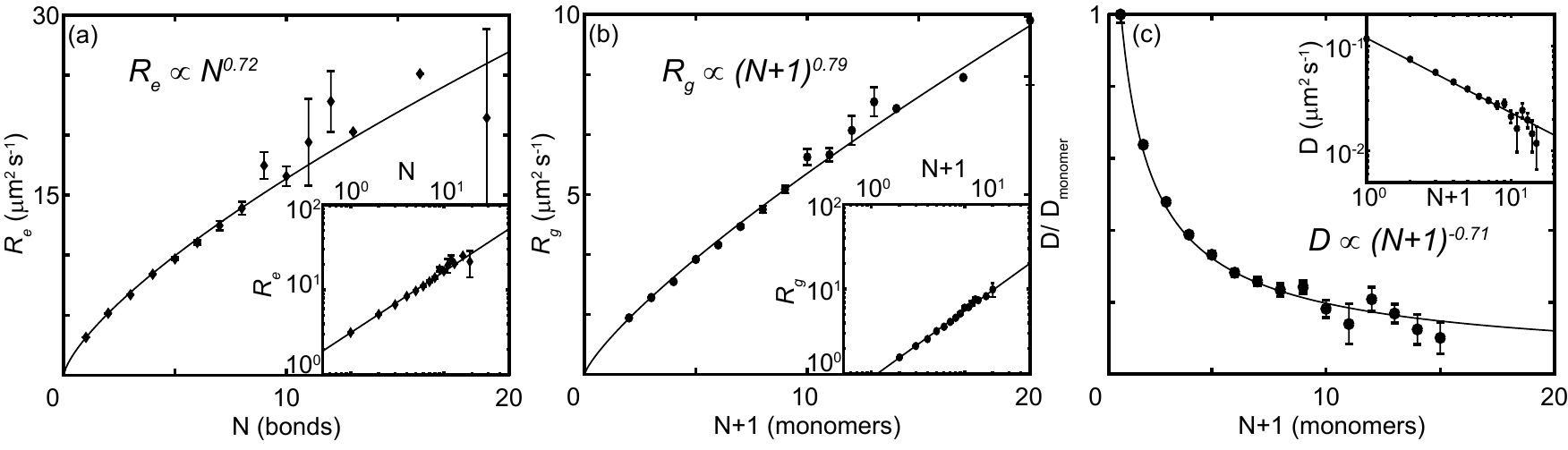}%
     \caption{Plots of the (a) $R_e$, (b) $R_g$, and (c) $D$, normalized by the diffusion of a single monomer ($0.12\,\mu$m$^2$s$^{-1})$, for linear emulsion polymers as a function of the number of bonds $N$, where $N+1$ is the number of droplets in the chain. Error bars are the standard error from measurements of many polymers and time points. Insets show the scaling relations on a log-log plot.  }
     \label{fig:polymer}
 \end{figure*}
  
Figure\,\ref{fig:joint}(b) shows that the measured $D_{rot}$ does not depend on the concentration $C_{DNA}$. 
For our 3$\mu$m particles, the average value of $D_{rot} = 0.008$\,rad$^2$\,s$^{-1}$ corresponds to a translational diffusion coefficient of $D_T = 0.07 \,\mu$m$^2$\,s$^{-1}$. 
This value is of the same order of magnitude as the measured diffusion coefficient of a single monomer, $D_{m} = 0.12 \pm 0.03\,\mu$m$^2$\,s$^{-1}$, and similar to previously reported values~\cite{perry2015two}. 
We conclude that the reorganization of bonds in space is dominated by the diffusion of the droplet in the fluid, rather than the DNA patch sliding around on the droplet surface.
The uniform distribution in Fig.\,\ref{fig:joint}(a) also implies that the Kuhn length $b$ of our polymers is the distance between two bound droplets, i.e. the diameter.

Considering only linear colloidomers from all $C_{DNA}$ conditions, we characterize the structure of $\approx28000$ chains of lengths ranging from dimers to twentymers. Figures\,\ref{fig:polymer}(a,b) show the sublinear growth of the root-mean-square end-to-end distance, $R_e =\sqrt{\left< \mathbf{R}^2 \right>}$, and the radius of gyration, $R_g$, where
  \begin{equation}
 R_g^2=\frac{1}{(N+1)^2}  \sum_{0\leq i < j \leq N} \langle r^2_{ij} \rangle 
 \end{equation}
as a function of the number of bonds $N$ and the number of droplets $N+1$ in a colloidomer, respectively. 
Each point is an ensemble average (in time and space) over thousands of polymers, ensuring ergodic sampling of the configurational space. 
The error bar increases with chain length because there are fewer long chains. 

 

We find that both size parameters scale in good agreement with the Flory theory of a self-avoiding polymer, which happens to be exact in 2D, as $R_{e} = b N^\nu$ and $R_g = \alpha bN^\nu$, where $\alpha$ is a unitless constant, $b$ corresponds to the Kuhn length, and $\nu=3/4$.
Specifically, fits to the data yield $b = 1.0\pm 0.03$ diameter, $\alpha = 0.3\pm 0.02$, and exponents $\nu=0.72\pm0.03$ and $0.79\pm0.02$ for $R_{e}$ and $R_g$, respectively. 
It is surprising that short chains already obey the scaling law which is supposed to apply in the asymptotic $N \to \infty$ limit only. 
One explanation is that the Ginzburg parameter is at its upper limit~\cite{Kremer1981}. 
The Ginzburg parameter for a polymer chain in $d$ dimensions is given by the ratio $v/b^d$, where $v$ is the excluded volume of a monomer and $b$ is the segment length \cite{RedBook}.
In our case, the segment length is very nearly equal to the diameter of the excluded area. 
Accordingly, unlike regular polymers and particularly semi-flexible ones (e.g. dsDNA), our colloidomers exhibit no regime of polymer chain length with marginal self-avoidance. 
Therefore, the large $N$ scaling is visible beginning with the shortest of chains and agrees with scaling laws in molecular polymers, including DNA~\cite{Maier1999} in the large $N$ limit.

Next, we measure the dynamics of the colloidomer through the diffusion of its center of mass. 
Our system consists of micron-sized droplets rearranging in a fluid, which implies hydrodynamic correlations between the monomers. 
The diffusion of a coil is described by the Zimm theory, which predicts that the diffusion coefficient scales as $D\propto (N+1)^{-\nu}$. 
Since the chain does not move as a ball, but rather as a pancake in 2D, the exponent is predicted to be $\nu=3/4$, just like in the cases of $R_e$ and $R_g$, as long as the surrounding fluid is in 3D. Our data in Fig.\,\ref{fig:polymer}\,(c) is best fit with $\nu=0.71\pm0.11$.

In conclusion, this work shows that the mobility of DNA on the droplet surface allows for the self-assembly of colloidal, fully flexible polymers. 
The DNA-binder concentration controls the droplet valence, such that we obtain colloidomers with chain lengths whose distribution agrees with random attachment. 
Their bulk synthesis allows us to demonstrate that they are freely-jointed, and that their size and diffusion coefficient scale with the chain length in agreement with a self-avoiding polymer.

In the future, colloidomers will serve to build both jammed~\cite{Brown2012} and complex ordered phases, such as close-packed spirals, predicted by numerical simulations~\cite{Nguyen2015}. 
Even further, these colloidomers open the platform for the self-assembly of complex three dimensional materials or soft structures via a biomimetic folding approach~\cite{zeravcic2014size}. 
The higher valence emulsions assemble into fibrous colloidal gels, mimicking the structure of cytoskeletal actin or microtubule networks. 
Just as their structure is continuously reconfigured by the cell, the reversible nature of the droplet-droplet bond paves the path towards responsive soft matter.

\begin{acknowledgments}
The authors gratefully acknowledge insights by David Pine and Sascha Hilgenfeldt, as well as useful conversations with Etienne Ducrot and Matan Yah Ben Zion. This work was supported by the Materials Research Science and Engineering Center (MRSEC) program of the National Science Foundation under Grant No. NSF DMR-1420073, No. NSF PHY17-48958 and No. NSF DMR-1710163. 
H.-C. was partially supported by Department of Energy grant DE-SC0012296 and the Alfred P. Sloan Foundation. 
FS is grateful to the CSMR at NYU for hospitality.
\end{acknowledgments}

\bibliographystyle{apsrev4-1} 
\bibliography{bibliography.bib}



\end{document}


\beginsupplement

\preprint{APS/123-QED}

\title{Supplementary Material: Colloidal Polymers}
\author{Angus McMullen}
 \affiliation{Physics Department, New York University, New York, New York 10003, USA}
\author{Miranda Holmes-Cerfon}
 \affiliation{Courant Institute of Mathematical Science, New York University, New York, New York 10012, USA}
  \author{Francesco Sciortino}
 \affiliation{Physics Department, Sapienza Universit\a' di Roma, Rome, Italy}
 \author{Alexander Y. Grosberg}
 \affiliation{Physics Department, New York University, New York, New York 10003, USA}
\author{Jasna Brujic}
 \affiliation{Physics Department, New York University, New York, New York 10003, USA}

\date{\today}

\begin{abstract}
\end{abstract}

\maketitle

\section{Experimental Methods}
We synthesized monodisperse PDMS droplets via a method adapted from ref ~\cite{Elbers2015}. 
Briefly, DMDES monomer, ranging from 1\% v/v to 20\% v/v, was dissolved in deionized water via vortexing. 
The droplet size was tuned via the amount of monomer added. 
Once the solution was clear, and the monomer fully dissolved, enough ammonia was added to bring the v/v\% of ammonia to 20\%. 
The solution immediately turned cloudy, and sodium dodecyl sulfate (SDS) was added to bring the concentration to 1 mM. 
This stabilized the droplets during growth, decreasing polydispersity. 
The sample was then left on a rotator for 24 hours, and then dialyzed against 1 mM SDS to remove any unreacted monomer, ammonia, and ethanol from  solution. 
Droplets were then stored in 10 mM SDS until ready for use. 
After the droplets were synthesized, they were incubated with a variable amount of binding DNA which had already been attached to a distearoyl-sn-glycero-3-phosphoethanolamine-N-[dibenzocyclooctyl(polyethylene glycol)-2000]  (DSPE-PEG2000-DBCO, Avanti) molecule. 
The hydrophobic part of the lipid anchored the DNA on the surface of the droplet. 
To give the DNA extra stability, and ensuring that it would not migrate during washing or incubation with droplets of other flavors of DNA, we anchored both the DNA strand containing the 20 mer sticky end, as well as a strand that was complementary to the 50 base pseodo random spacer prior to the sticky end.
We used the following DNA strands as our binding pair: 5$\prime$-Azide-Cy3-A GCA TTA CTT TCC GTC CCG AGA GAC CTA ACT GAC ACG CTT CCC ATC GCT A      
GA GTT CAC AAG AGT TCA CAA-3$\prime$ and 
5$\prime$-Azide-Cy5-A GCA TTA CTT TCC GTC CCG AGA GAC CTA ACT GAC ACG CTT CCC ATC GCT ATT GTG AAC TCT TGT GAA CTC-3$\prime$. The five prime end of each strand was terminated with an azide group, and each strand was labeled with a Cy3 or Cy5 fluorophore, respectively. The last 20 bases on each strand constituted our binding sequence, while the first 50 bases formed our spacer. This spacer was complexed with the following complementary strand: 5$\prime$-TAG CGA TGG GAA GCG TGT CAG TTA GGT CTC TCG GGA CGG AAA GTA ATG CT-Azide-3$\prime$. Both binding sequences and complementary spacer strand were reacted with 1,2-distearoyl-sn-glycero-3-phosphoethanolamine-N-[dibenzocyclooctyl(polyethylene glycol)-2000]  (DSPE-PEG2000-DBCO, Avanti) through a copper free click reaction. The DSPE anchored the DNA strands to the surface of the droplet. 
We washed our samples in 50\,mM NaCl with 10 mM Tris pH 8 and 1 mM EDTA. Our assembly buffer was 20 mM MgCl$_2$ with 5 mM Tris pH8, and 0.1\% F68 pluronic surfactant.
We tracked our droplets using the MATLAB implementation of the Crocker Grier tracking algorithm \cite{Crocker1996}. 
Adhesion patches were not always visible between two bound droplets during movie acquisition because of the short exposure time. In order to determine if two droplets were bound, therefore, we tracked them over time. If two droplets remained connected over a given time period (typically one minute), we consider them bound. 
Fluorescence images in Fig.\.(2) in the main text were taken with droplets prepped in a different fashion. 
Instead of connecting the DNA through a copper-free click reaction to the DSPE lipids, we connected the DNA to the DSPE lipids via a biotin-streptavidin sandwich as described in refs. \cite{Feng2014,Zhang2017}. 
This method allowed for detailed fluorescence imaging, but lead to many experimental issues including difficulties with temperature reversability and nonspecific binding especially at long observation times. 

Temperature cycling was done on a custom built heat stage over a long working distance air objective.
While the amount of F68 surfactant we used was well below the CMC at room temperature, the CMC of F68 decreases as the temperature increases. 
The faster aggregation kinetics at high temperatures shown in the main text is due to an additional depletion effect from the F68 surfactant that only occurs above $T \approx 45^\circ$ degrees C. 

\section{Random Branching Model}

We model the size distribution of clusters assuming that (a) monomers attach independently of each other and uniformly on the set of available bonding sites, (b) clusters grow in a bath of monomers with constant valency distribution; (c) each cluster is saturated, i.e. there are no more open bond sites to which a monomer can attach, and (d) the clusters cannot form loops, only linear or branch-like structures. 

The model's input is the distribution of valencies. Let $p_i$ be the probability that a randomly-chosen monomer has valence $i$. For example, if the possible monomer valencies are 1, 2, or 3, then the valence distribution is characterized by parameters $(p_1,p_2,p_3)$ with $p_1+p_2+p_3=1$. 
 
 Consider a growing cluster and label the particles in the order they attach to the cluster.
Each particle that attaches changes the total valence (number of open bonding sites) of the cluster: if a particle with valence $\nu$ attaches, the change in the total valence is $\nu-1$. 
Let $X_i$ be the net change in valence of the cluster after the $i$th particle attaches; it has probability distribution $P(X_i=k-1) = p_k$. Let $V_i$ be the total valence of a cluster of size $i$, which for $i\geq 2$ is given by
\begin{equation}
V_i = V_1 + X_2 + \cdots + X_i\,.
\end{equation}
By the assumption that particles attach independently of each other and of their history, $V_1,V_2,\ldots$ forms a time-homogeneous Markov chain, with transition probabilities 
$P(V_{i+1}=k|V_i=j)  = p_{k-j+1}$.

Let $L$ be the length of a saturated cluster. We wish to compute its probability distribution,  $P(L=n)$, $n=2,3,\ldots$ (ignoring the monomers remaining in the solution.)

Notice that a cluster is saturated when its valence hits 0. 
The length of such a cluster is the first value of $i$ for which $V_i=0$. Solving for this distribution is a first-passage-time problem and may be done using using standard methods for time-homogeneous Markov chains \cite{Grimmett2001}. Let 
\begin{equation}
L_{n,i} = P(L=n|V_1=i)
\end{equation}
be the probability the length is $n$, given the first particle has valence $i$. Then $L_{n,i}$ solves a linear, recursive equation
\begin{equation}\label{L}
L_{n,i} = p_1L_{n-1,i-1} + p_2L_{n-1,i} + \cdots + p_kL_{n-1,i+k-2}+ \cdots.
\end{equation}
The initial and boundary conditions are $L_{1,k}=\delta_{0k}$ ($\delta_{0k}=1$ if $k=0$, $\delta_{0k}=0$ otherwise),  $L_{k,0}=0$ ($k\geq 2$).

Given $L_{n-1,\cdot}$ for some $n$, we can use \eqref{L} 
to solve for the values of $L_{n,\cdot}$, and then compute the length distribution by averaging over the initial conditions with the correct probabilities: 
\begin{multline}
P(L=n) = \sum_k P(L=n|V_1=k)P(V_1=k) \\= \sum_k L_{n,k}p_k\,.
\end{multline}

While this computation can in principle be done analytically, for example using a symbolic software package, the formulas it gives are complicated and unhelpful so we generally solved the model numerically up to a desired value of $n$. The exception is when particles have maximum valence 2, in which case one can easily obtain the analytic formula $P(L=n)  = (n-1)p_1^2(1-p_1)^{n-2}$.

Note that despite the monomer addition assumption, the model captures also cluster-cluster aggregation. If a cluster grows by two smaller clusters attaching together, we may still assign a canonical order to the particles, by labeling the particles in the order they formed bonds, and shifting the numbers so the growing cluster is always connected. For example: suppose there are 5 particles with labels A,B,C,D,E, and they form bonds in the following order: B-C, D-A, C-E, D-E. Then we would assign labels B=1,C=2,E=3,D=4,A=5.

We remark that our model is closely related to the branching processes studied in mathematical probability theory \cite{Grimmett2001}. One can show that the generating function $H(s)$ for the total cluster size must satisfy the equation (\cite{Grimmett2001})
\[
H(s) = sG(H(s))\,,
\]
where $G(s)$ is the generating function for $X_i$. 
One can solve for $H(s)$ analytically provided the maximum valence is 3 or less. However, obtaining the distribution of lengths $(P(L=n))_{n=2}^\infty$ from the generating function appears just as complicated  as solving the recursive equation \eqref{L}.

\section{Angular Diffusion}
Angular diffusion in 2D is described by
\begin{equation} \frac{\partial P}{\partial t} = D_{\mathrm{rot}} \frac{\partial^2 P}{\partial \theta^2} + \delta(t) \delta(\theta) \ , \end{equation}
subject to boundary condition $P(-\pi) = P(\pi)$.  The formal solution reads 
\begin{equation}
P(\theta,t) = \frac{1}{2 \pi} + \frac{1}{\pi} \sum_{m=1}^{\infty} e^{-D_{\mathrm{rot}} m^2 t} \cos m \theta 
\end{equation}
which then implies 
\begin{equation}
\left< \cos \theta \right> = \int_{-\pi}^{\pi} P(\theta, t) \cos \theta d\theta =  e^{- D_{\mathrm{rot}} t}
\end{equation}
or $D_{\mathrm{rot}} = -\ln \left< \cos \theta \right> /t$.  

\section{Monte Carlo simulation}

We perform Monte Carlo (MC) simulations of a mixture of particles with prescribed valence, with the same distribution of patches and the same surface coverage as the one experimentally measured.  We model the patchy particles as hard disks of diameter $\sigma$ decorated with $n$ patches.  Patches are located on the surface of 
the particle in the most symmetric geometry ($2\pi/n$, with n=1,...5). As in the experiments,
for each $n$, half of the particles are considered of type $A$ and half of type $B$.
Pairs of particles of different types ($A$ and $B$) interact via the well known Kern-Frenkel~\cite{kern2003fluid} potential in two dimensions, e.g. an angularly modulated square well of depth $\epsilon$ and
angular width $\pm 25 \deg$.
The MC move is composed by a random translation by $0.15 \sigma$ and a random rotation of $0.3$ radiants. For each of the investigated concentrations of $DNA$ we perform both simulations in
thermodynamic equilibrium down to $k_BT/\epsilon=0.05$ as well as in
irreversible conditions, by performing a $T$ jump down to $k_BT/\epsilon=0.001$, starting from a configuration of not-bonded particles. In the latter case, no bond-breaking 
events are observed in the course of the simulation.  We simulate a system of
2000 particles and average the results over 200 different realizations. 
In both simulation methods (equilibrium and irreversible aggregation) in the final configuration more than $99 \%$ of all possible bonds are formed.  
The two methods generate  very similar cluster size distributions in the final state,
with only slightly larger clusters in the equilibrium simulations. 
The resulting irreversible cluster size distributions are compared with the experimental ones in Fig.~3
of the manuscript.

\section{Polymer Diffusion}

The self diffusion of each polymer is measured by first determining the center of mass position of each polymer over time. 
We then measure the mean squared displacements (MSDs) of the center of mass position as a function of the time lag $\delta t$.
Polymers with any monomers non-specifically stuck to the glass are excluded. 
To measure $D$ for a given polymer, we fit a line to points on the MSD vs $\delta t$ between 0.5 seconds and 3 minutes.

\begin{figure}[!hbt]
   \centering
   \includegraphics[width = \columnwidth]{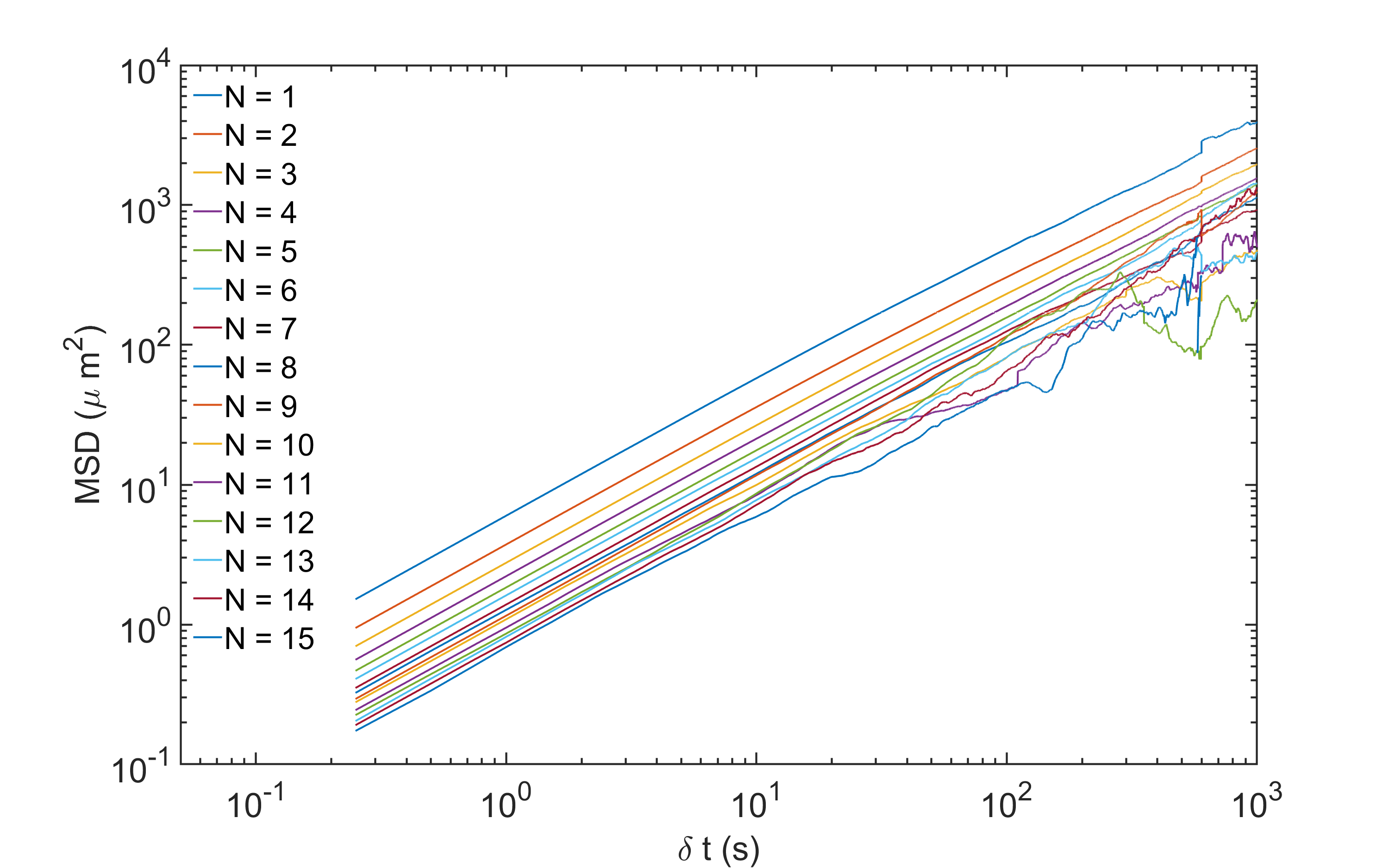} 
   \caption{Mean squared displacements as a function of $\delta t$ for colloidal polymers of different lengths. $N$ is the number of bonds. The MSD shown is compiled from all observed polymers of each respective length, filtering out polymers that stuck. }
   \label{fig:polymerdiffusion}
\end{figure}

\section{Assembly Rates}

\begin{figure}[!hbt]
   \centering
   \includegraphics{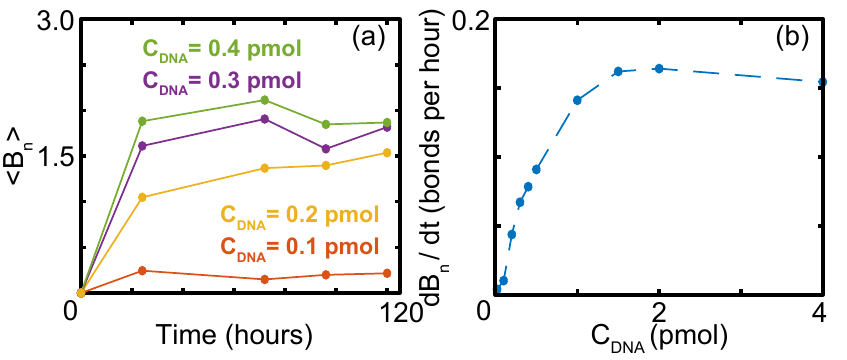} 
   \caption{(a) The average bond number for a given loading of DNA as a function of incubation time. (b) The average rate at which bonds are added in the first $\approx$ 24 hours as a function of the DNA loading.}
   \label{fig:rates}
\end{figure}

Figure\,\ref{fig:rates} (a) shows the average bond number of a sample of droplets for $C_{DNA} = 0.1, 0.2, 0.3,$ and 0.4 pmol DNA loading as a function of incubation time. 
The bond number increases sharply after the first 24 hours and then is approximately stable, with some slow growth in the 0.2 pmol condition. 
Figure\,\ref{fig:rates} (b) shows the approximate rate at which bonds are added per hour, over a 24 hour period. 
The rate is highly dependent on DNA loading until approximately $C_{DNA} = 1.75$ pmol, at which point it stabilizes likely because bond number is limited in those conditions by sterics and not the availability of DNA. 

\section{Supplemental Movie Descriptions}

The Supplemental Movie 1 shows a freely jointed colloidomer twenty monomers long diffusing. 
This movie is speed up by a factor of five.
Supplemental Movie 2 shows a freely jointed colloidomer of three monomers as well as overlayed schematic demonstrating the angle subtended by the two end caps of the trimer. 
This movie is shown in real time.

\bibliographystyle{apsrev4-1} 
\bibliography{SMbib.bib}